\documentclass[a4paper,11pt,preprint]{article}

\usepackage{jinstpub} 

\usepackage[left]{lineno}
\subheader{{\rm Report number: KYUSHU-RCAPP-2020-02}}
\title{\boldmath Measurement of $\gamma$ rays from $^6$LiF tile as an inner wall of a neutron-decay detector}

\author[a,1]{J. Koga\note{Corresponding author.}}
\author[b,2]{, S. Ieki\note{Present address : Research Center for Neutrino Science, Tohoku University, Sendai, Miyagi, 980-8578, Japan}}
\author[c]{, A. Kimura}
\author[d]{, M. Kitaguchi}
\author[e,3]{, R. Kitahara\note{Present address : Sumitomo Heavy Industries, Ltd., 19 Natsushima-cho, Yokosuka-shi, Kanagawa 237-8555, Japan}}
\author[f]{, K. Mishima} 
\author[b]{,\\ N. Nagakura}
\author[g,4]{, T. Okudaira\note{Present address : Japan Atomic Energy Agency, Tokai 319-1195, Japan}}
\author[h]{, H. Otono}
\author[g]{, H. M. Shimizu}
\author[a,5]{, N. Sumi\note{Present address : Cryogenics Science Center, KEK, Tsukuba, 305-0801, Japan}}
\author[a]{, S. Takada} 
\author[a,6]{,\\ T. Tomita\note{Present address : Tokyo Tatemono Co.,Ltd., Chuo-ku, Tokyo 103-8285, Japan}}
\author[b]{, T. Yamada}
\author[h]{, and T. Yoshioka}

\affiliation[a]{Department of Physics, Kyushu University, Fukuoka 819-0395, Japan}
\affiliation[b]{Department of Physics, The University of Tokyo, Bunkyo-ku, Tokyo, 113-0033, Japan}
\affiliation[c]{Japan Atomic Energy Agency, Tokai 319-1195, Japan}
\affiliation[d]{Kobayashi-Maskawa Institute, Nagoya University, Nagoya 464-8602, Japan}
\affiliation[e]{Department of Physics, Kyoto University, Kyoto 606-8502, Japan}
\affiliation[f]{High Energy Accelerator Research Organization, Tokai 319-1109, Japan}
\affiliation[g]{Department of Physics, Nagoya University, Nagoya 464-8602, Japan}
\affiliation[h]{Research Center for Advanced Particle Physics, Kyushu University, Fukuoka 819-0395, Japan}

\emailAdd{j.koga@epp.phys.kyushu-u.ac.jp}

\abstract{A neutron lifetime measurement conducted at the Japan Proton Accelerator Research Complex (J-PARC) is counting the number of electrons from neutron decays with a time projection chamber (TPC). 
The $\gamma$ rays produced in the TPC cause irreducible background events.
To achieve the precise measurement, the inner walls of the TPC consist of $^6$Li-enriched lithium-fluoride ($^6$LiF) tiles to suppress the amount of $\gamma$ rays.
In order to estimate the amount of $\gamma$ rays from the $^{6}{\rm LiF}$ tile, prompt gamma ray analysis (PGA) measurements were performed using germanium detectors.
We reconstructed the measured $\gamma$-ray energy spectrum using a Monte Carlo simulation with the stripping method. 
Comparing the measured spectrum with a simulated one, the number of $\gamma$ rays emitted from the$^{6}{\rm LiF}$ tile was $(2.3^{+0.7}_{-0.3}) \times 10^{-4}$ per incident neutron. 
This is $1.4^{+0.5}_{-0.2}$ times the value assumed for a mole fraction of the $^{6}{\rm LiF}$ tile.
We concluded that the amount of $\gamma$ rays produced from the $^{6}{\rm LiF}$ tile is not more twice the originally assumed value.}


\begin{document}
\maketitle
\flushbottom

\section{Introduction}
\label{sec:intro}

Accurate estimation and suppression of background events in experimental particle physics and nuclear physics are very important to realize precise measurements. 
$^6$Li-enriched lithium-fluoride ($^6$LiF) is often used as neutron shields to suppress background events because $^6$Li has a large neutron-absorption cross section and low $\gamma$-ray emission probability after absorbing neutrons.

Neutron lifetime measurements should be precise, because it is an important parameter for the $V_{ud}$ term in the Cabibbo-Kobayashi-Maskawa matrix and it is also used to verify the Big Bang nucleosynthesis. 
Two methods have historically been used in neutron lifetime measurements: counting the number of protons from neutron decays \cite{a,b}, and counting the number of surviving neutrons after a certain stored time \cite{c,d,e,f,g}.
However, there is a significant deviation ($8.4 \pm 2.1~{\rm sec}$ or 4.0$\sigma$) between the results from these two methods. Therefore, more accurate measurement methods are necessary.

An experiment being conducted at the Japan Proton Accelerator Research Complex (J-PARC) is the first experiment using a pulsed neutron beam to measure the neutron lifetime. 
In this experiment, the number of electrons from neutron decays is counted with a time projection chamber (TPC) filled with an operation gas ($^4$He, ${\rm CO_2}$, and a few ppm of $^3$He gas) \cite{h,NeutronLifetimeJPARC}. 
This TPC can observe neutron decay events by detecting the drifting ionized particles with a multi-wire proportional chamber. 
Neutrons are often scattered by the operation gas and hit the inner walls of the TPC. 
If $\gamma$ rays produced by capturing neutrons hit the inner materials, electrons are emitted from the hit positions. 
These electrons cause irreducible background events and a large uncertainty in subtraction \cite{NeutronLifetimeJPARC}, and thus, the inner walls of the TPC consist of $^6$LiF tiles to suppress the amounts of emitted $\gamma$ rays.

However, one must pay attention to the amounts of $\gamma$ rays from the $^6$LiF tiles per incident neutron to achieve precise measurement. 
For this purpose, we conducted prompt gamma ray analysis (PGA) measurements of the $^6$LiF tile with a pulsed neutron beam and germanium detectors.
The obtained spectrum was reconstructed using the stripping method \cite{Stripping} with a Monte Carlo simulation.
As a result, we estimated the intensity and energy of $\gamma$ rays emitted from the inner walls of the TPC per neutron.

This paper is organized as follows: the detailed properties of the $^6$LiF tiles and the equipment in neutron-irradiation experiments are described in section~\ref{sec:experiment}. Then, we report measurements and analysis result in section~\ref{sec:ana}. Finally, we summarize this work in section~\ref{sec:summary}.

\section{Experimental setup} \label{sec:experiment}

\subsection{Properties of the $^6$LiF tile} \label{subsec:6Li}
\begin{figure}[htbp]
\centering
\begin{tabular}{c}
	\begin{minipage}{0.47\hsize}
	\centering
	\vspace{5mm}
	\includegraphics[width=.7\textwidth,bb = 0 0 581 708,angle=-90]{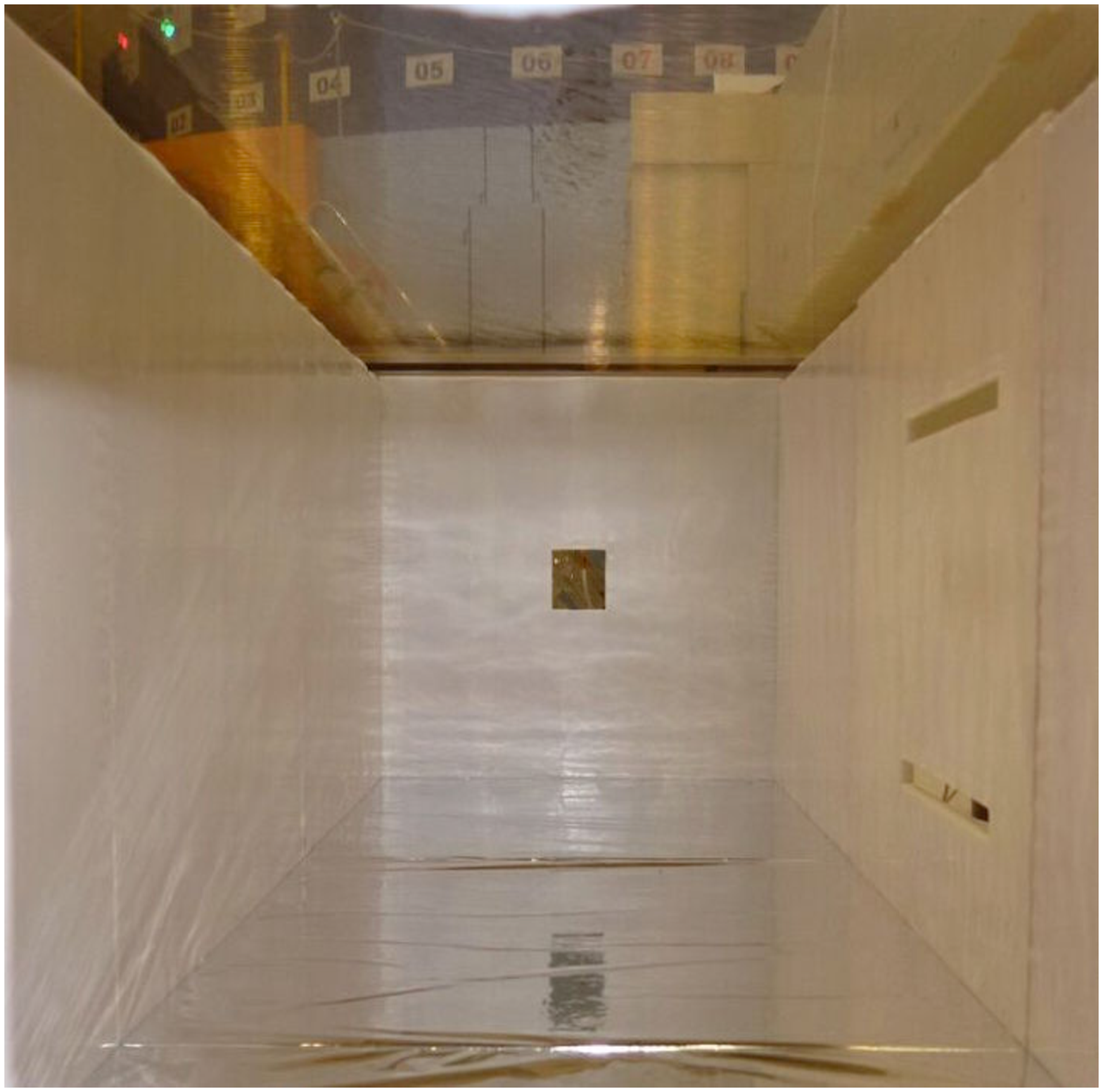}
	\caption{\label{fig:TPCphoto} Picture of the inside of the TPC taken from the downstream. The inner wall of the TPC consists of the $^6$LiF tiles.}
	\end{minipage}
	\begin{minipage}{0.06\hsize}
	\hspace{2mm}
	\end{minipage}
	\begin{minipage}{0.47\hsize}
	\centering
	\includegraphics[width=.7\textwidth,bb = 0 0 581 715]{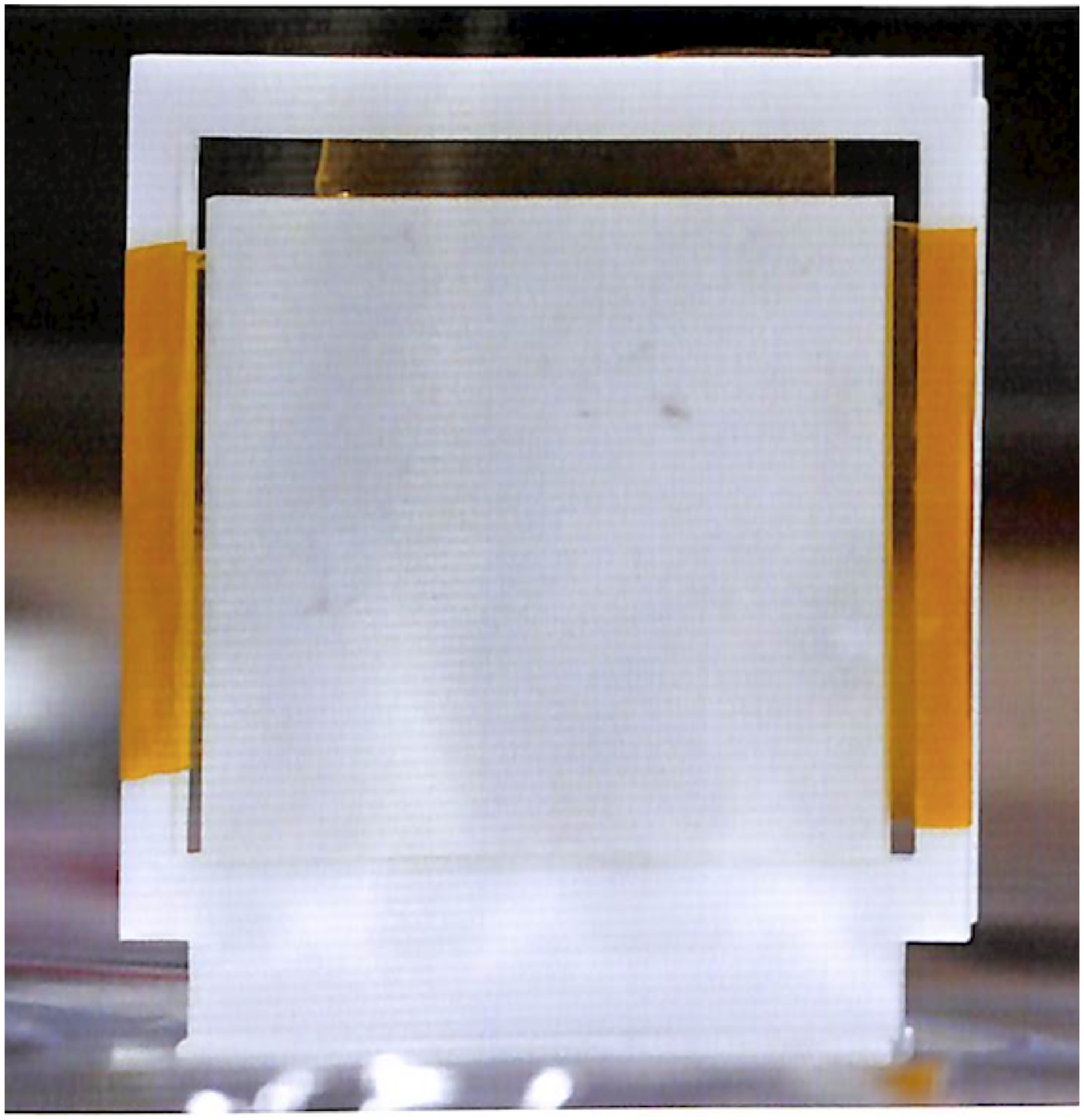}
	\caption{\label{fig:Sample_Li} Picture of the $^6$LiF tile with size $40 \times 40 \times 5~{\rm mm}^3$. This tile is used in a neutron irradiation experiment described in section~\ref{subsec:Collected data}.}
	\end{minipage}
\end{tabular}		
\end{figure}

Figure~\ref{fig:TPCphoto} is a picture of the inside of the TPC taken from the downstream. 
The inner wall of the TPC consists of $^6$LiF tile made by sintering $^6$LiF powder at 30 wt\% with poly-tetra-fluoro-ethylene (PTFE) at 70 wt\%. 
The enrichment of $^6$Li in this LiF powder is 95\%, and thus, neutron capture cross-section with $\gamma$ rays of the $^6$LiF tiles is expected to be small.
Table~\ref{tab:LiF} shows the constituent elements and properties of the $^6$LiF tile. 
Impurities in the $^6$LiF tile were inspected by X-ray fluorescence analysis (XRF). 
The total amount of impurities was about 0.1\%, and the detailed result is described in appendix~\ref{app:impurity}.
The thickness of the inner wall is 5~mm and an effective thickness $L_{\rm eff}$ is defined as 
\begin{equation}
L_{\rm eff} = \frac{1}{\sum_{i} \rho_i \sigma_{{\rm tot}, i}} \, ,
\end{equation}
where $\rho_i$ and $\sigma_{{\rm tot}, i}$ are the number density and the total cross section for each element, respectively.
$L_{\rm eff}$ is $6.5 \times 10^{-2}$~cm for 2200~m/s neutrons, which was calculated using values in table~\ref{tab:LiF}.
Figure~\ref{fig:Sample_Li} shows the tile with size $40 \times 40 \times 5~{\rm mm}^3$. 
This tile was used for the PGA measurement, which is described in section~\ref{subsec:Collected data}. 

\begin{table}[htbp]
\centering
\caption{\label{tab:LiF} Constituent elements and properties of the $^6$LiF tile. The values of cross sections have been adopted from ref.~\cite{Mughabghab}.}
\smallskip
\begin{tabular}{|c|c|c|c|c|}
\hline
Element & \begin{tabular}{c} Mole \\fraction \end{tabular} & \begin{tabular}{c} Number density \\{[\#/cm$^3$]} \end{tabular} & \begin{tabular}{c} Capture cross section\\ {[mbarn]} \end{tabular}& \begin{tabular}{c} Total cross section \\ {[barn]} \end{tabular} \\
\hline
$^6$Li & 0.173 & $1.60 \times 10^{22}$ & 38.5 & 941\\
$^7$Li & 0.009 & $8.30 \times 10^{20}$ & 45.4 & 1.02\\
C & 0.212 & $1.96 \times 10^{22}$ & 3.50 & 4.74\\
F & 0.606 & $5.59 \times 10^{22}$ & 9.51 & 3.65\\
\hline
\end{tabular}
\end{table}

\subsection{Experimental equipment} 
\begin{figure}[htbp]
\centering
\includegraphics[width=.7\textwidth]{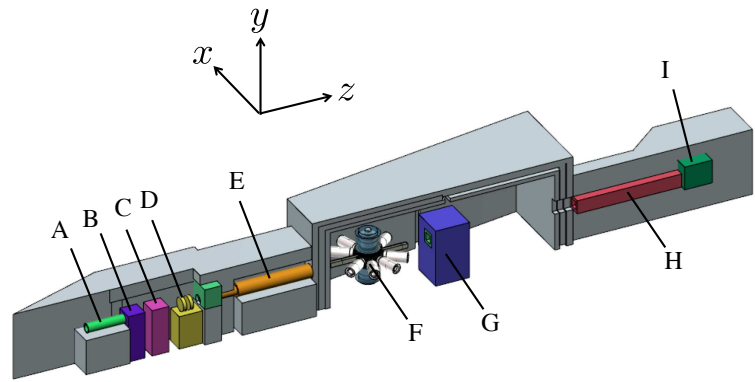}
\caption{\label{fig:beamline} Schematic illustration of the beamline 04. (A) Collimator, (B) T0-chopper, (C) Neutron filter, (D) Disk choppers, (E) Collimator, (F) Germanium detector assembly, (G) Collimator, (H) Boron resin, and (I) Beam stopper (Iron) \cite{Takada}.}
\end{figure}

\begin{figure}[htbp]
\centering
\includegraphics[width=.6\textwidth, bb=37.5 15.5 600.5 500.5]{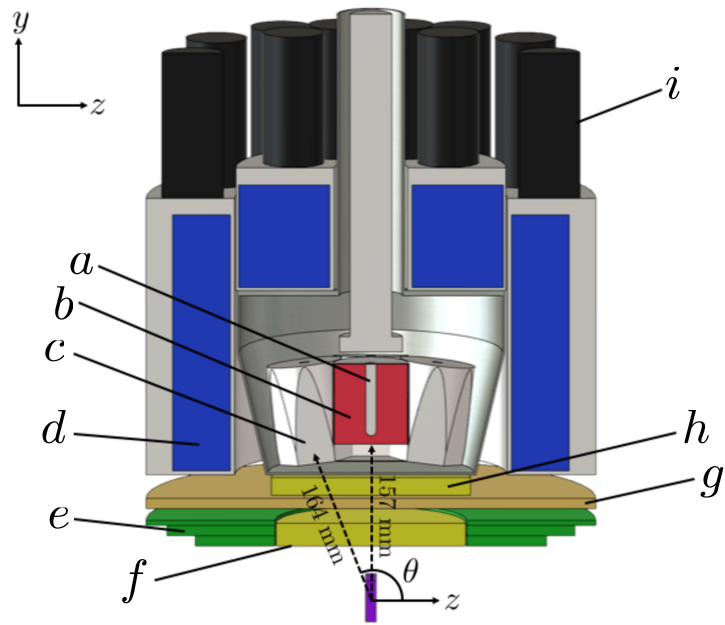}
\caption{\label{fig:Ge} Schematic of a detector unit containing of seven germanium crystals. (a) Electode, (b) Germanium crystal, (c) Aluminum case, (d) BGO crystal, (e) $\gamma$-ray shield (Pb collimator), (f) Neutron shield-1 (22.3 mm LiH), (g) Neutron shield-2 (5 mm LiF), (h) Neutron shield-3 (17.3 mm LiH), and (i) Photomultiplier tube for BGO crystal \cite{Takada}.}
\end{figure}

A germanium detector assembly is installed at the beamline 04 (ANNRI) of the Material and Life Science Facility (MLF) in J-PARC.
We can measure neutron capture cross-section accurately by detecting prompt $\gamma$ rays emitted from a target via (n,$\gamma$) reactions \cite{Igashira,Kino,Kimura}. 
Figure~\ref{fig:beamline} shows a schematic view of the beamline 04. 
Neutrons are produced by nuclear spallation reactions by irradiating a mercury target with a proton beam.
The neutrons are slowed down by a moderator and the pulsed neutron beam with a repetition rate of $25~{\rm Hz}$ passes through the neutron filter (C in figure~\ref{fig:beamline}), disk choppers (D in figure~\ref{fig:beamline}), and collimators (A and E in figure~\ref{fig:beamline}). 
The target and the germanium detector assembly (F in figure~\ref{fig:beamline}) are located at $21.5~{\rm m}$ downstream of the moderator surface. 
Disk choppers are used to avoid frame overlap by eliminating cold neutrons. 
Several neutron filters are inserted to adjust the beam intensity in the energy region of interest.
Seven germanium crystals form a detector unit, as shown in figure~\ref{fig:Ge}. 
These germanium detector units are covered with three neutron shields (f, g, and h in figure~\ref{fig:Ge}). 
Two units of germanium detector assembly are used in the measurement, and they are placed at the top and bottom of the target.
Usually, bismuth germanate (BGO) crystals (d in figure~\ref{fig:Ge}) are placed outside the germanium detectors and play a role to veto Compton scattering events, but they were not used in our measurements.

The output signal from each germanium crystal is processed independently. 
The output signals were fed into a module CAEN V1724 \cite{CAEN} that can record the pulse height digitized by an analog-to-digital converter (ADC), and the time interval between injection of the proton beam bunch and the detection time of the $\gamma$ ray emitted from the target.
This measured time information corresponds to the time-of-flight (TOF) of the neutrons from the moderator surface to the target.

\section{Measurement and analysis} \label{sec:ana}

\subsection{Measurement} \label{subsec:Collected data}
\begin{figure}[htbp]
\centering
	\begin{tabular}{c}
	\begin{minipage}{0.47\hsize}
	\centering	
	\includegraphics[width=\textwidth]{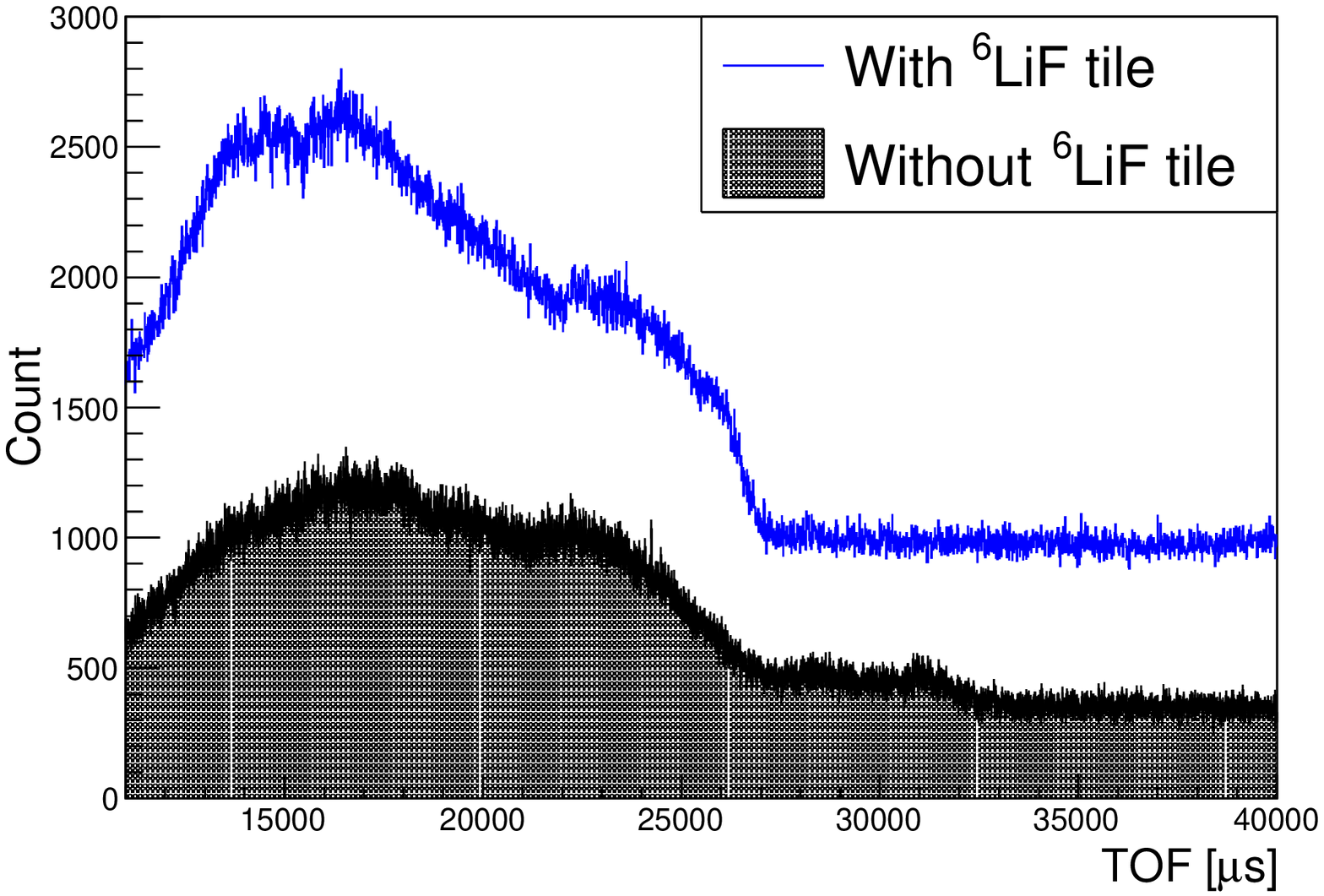}
	\caption{\label{fig:TOF} The spectra measured with the $^6$LiF tile (solid line) and without the $^6$LiF tile (dotted area).}
	\end{minipage}
	\begin{minipage}{0.06\hsize}
	\hspace{2mm}
	\end{minipage}
	\begin{minipage}{0.47\hsize}
	\centering
		\vspace{5mm} 
		\includegraphics[width=\textwidth]{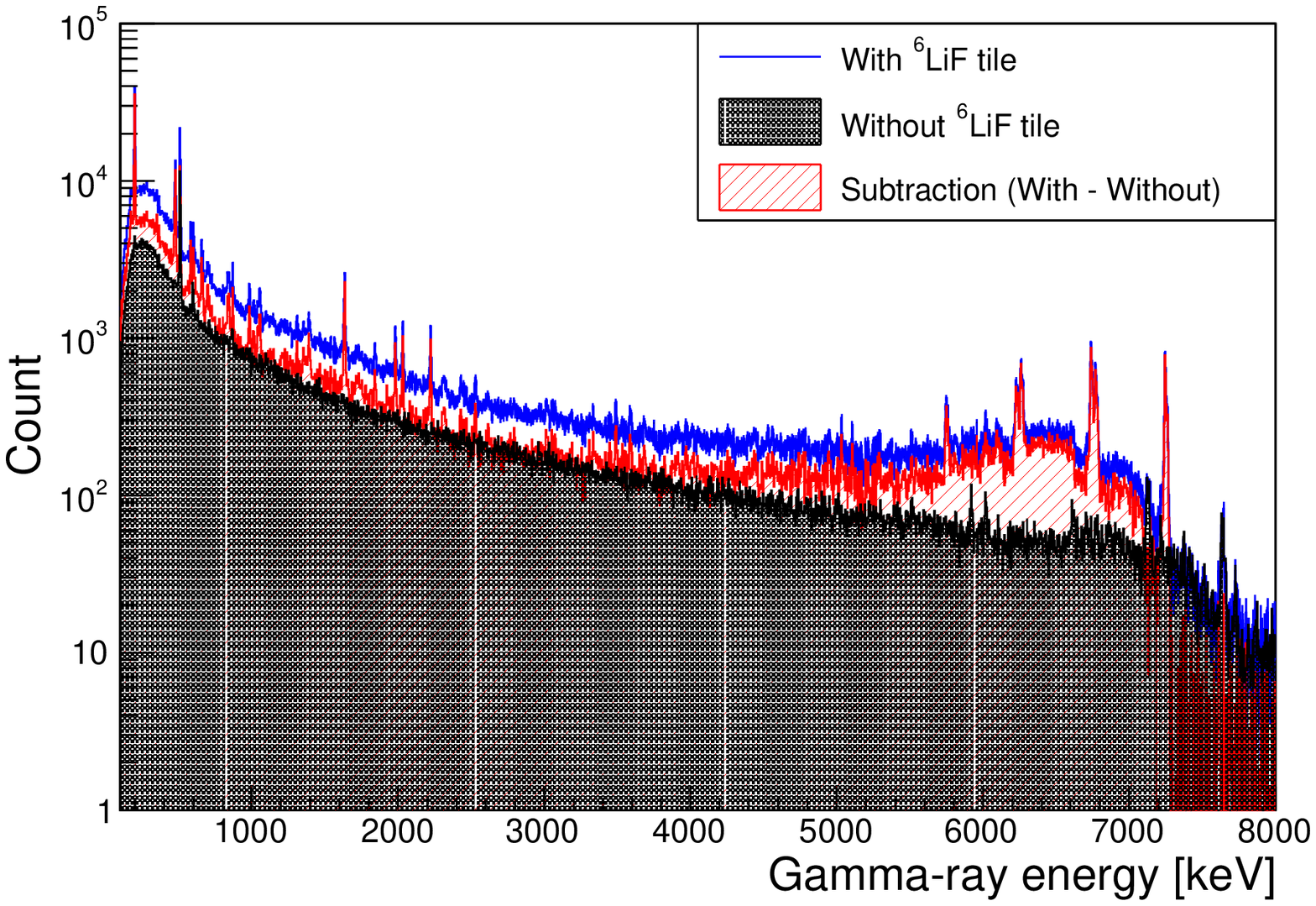}
		\caption{\label{fig:gE} The spectra measured with the $^6$LiF tile (solid line), without the $^6$LiF tile (dotted area), and their difference (shaded area).}
	\end{minipage}
	\end{tabular}
\end{figure}

The $^6$LiF tile was irradiated with the pulsed neutron beam with a time-averaged proton beam power of $200~{\rm kW}$ during the measurements. 
Measurements were performed with and without the $^6$LiF tile. The measurement time was for $35~{\rm minutes}$ and $42~{\rm minutes}$, respectively. 
The $\gamma$-ray energy calibration was performed on the basis of the intense peaks from the $^{27}$Al(n, $\gamma$) reactions.
In these measurements, the beam size on the target was adjusted as $22~{\rm mm}$ in diameter by the collimators. 
Cold neutrons that reach the target after about $27~{\rm ms}$ were blocked by the disk choppers, which is shown in figure~\ref{fig:TOF}.
Thus, the events after $27~{\rm ms}$ were not prompt $\gamma$ rays, but delayed $\gamma$ rays. 
The delayed $\gamma$ rays were removed by subtracting the spectrum after $27~{\rm ms}$.
The observed energy spectra of the prompt $\gamma$ rays are shown in figure~\ref{fig:gE}. 
The environmental background events were removed by subtracting the spectrum that was measured without the $^6$LiF tile from the one measured with the $^6$LiF tile. 
The spectrum of only the prompt $\gamma$ rays emitted from the $^6$LiF tile is also shown in figure~\ref{fig:gE}. 

\subsection{Analysis by stripping method and spectrum reconstruction} \label{subsec:ana}
The pulse heights of the output signals from the germanium detectors are not completely true $\gamma$-ray energies.
In some cases, $\gamma$ rays can not deposit their energy into the germanium crystals because of escape of photons generated by Compton scattering, trapping by lattice defects, etc. 
Therefore, the observed energies of $\gamma$ rays shift to lower values. 

The response function of each germanium detector is necessary in order to calculate the true $\gamma$-ray energy spectrum from the measurement.
In this study, the response function was calculated by a Monte Carlo simulation, and we calculated the intensity of each $\gamma$ ray peak emitted from $^6$LiF tile by the stripping method.
In the measured spectrum, the number of events in a certain energy region is the sum of the number of events of photoelectric absorptions and that of escapes of photons stemmed from $\gamma$ rays with higher energy.
The principle of the stripping method is as follows:
\begin{enumerate}
\item The number of events due to $\gamma$ rays with higher energy is subtract from that of events in a certain energy region using a simulated spectrum of monochromatic energy.
\item The number of events remaining after the above subtraction in the energy region is regarded as due to photoelectric absorption. The total amount of $\gamma$ rays emitted in all directions can be calculated using a detection efficiency and a geometrical acceptance.
\item These procedures are repeated from higher energies for the entire energy region. This is how the true $\gamma$-ray energy spectrum can be calculated.
\end{enumerate}

\begin{figure}[htbp]
\centering
	\begin{tabular}{c}
	\begin{minipage}{0.47\hsize}
	\centering	
	\includegraphics[width=\textwidth]{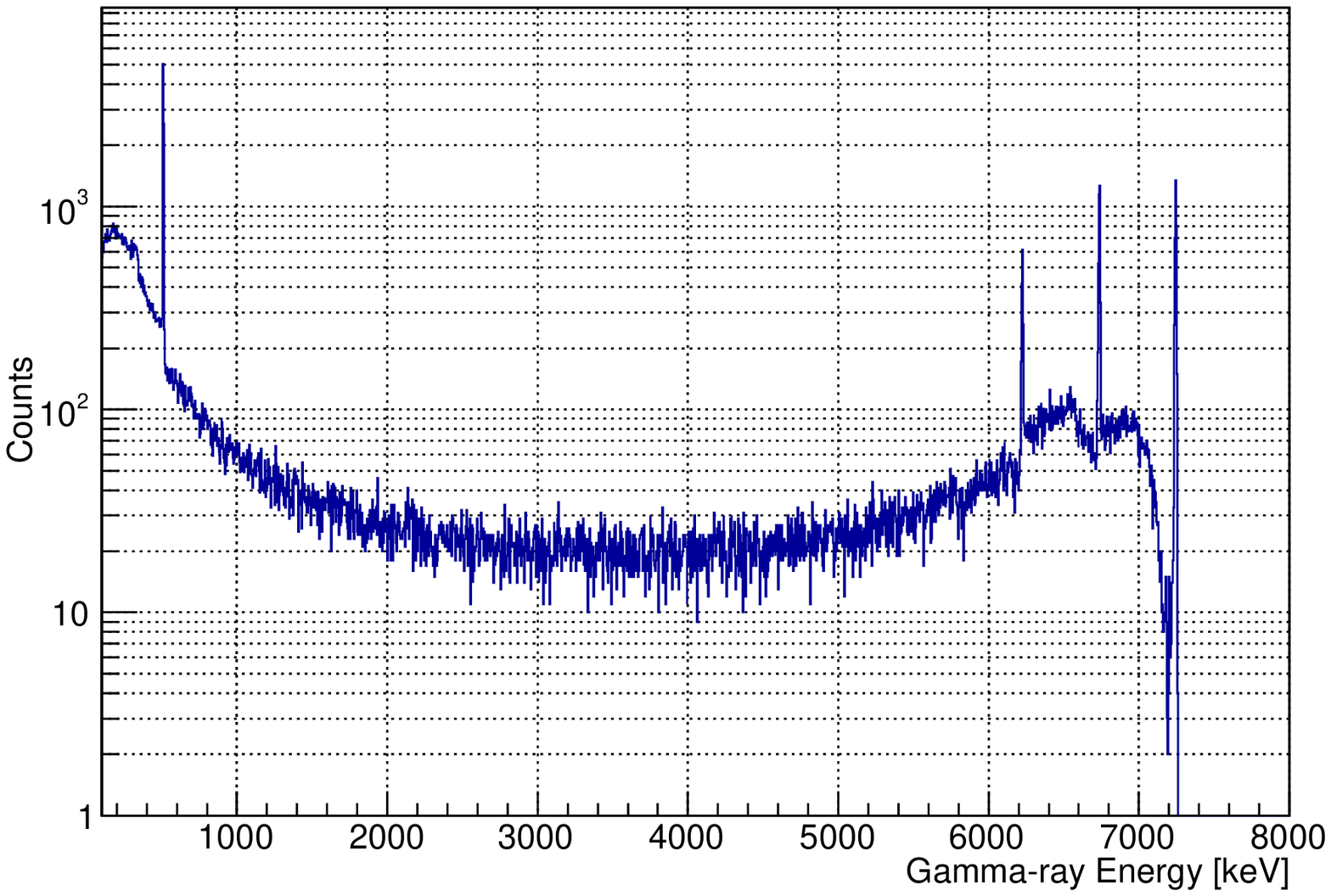}
	\caption{\label{fig:Simuhist} Simulated spectrum of monochromatic with energy 7245~keV.}
	\end{minipage}
	\begin{minipage}{0.06\hsize}
	\hspace{2mm}
	\end{minipage}
	\begin{minipage}{0.47\hsize}
	\centering
		\includegraphics[width=\textwidth]{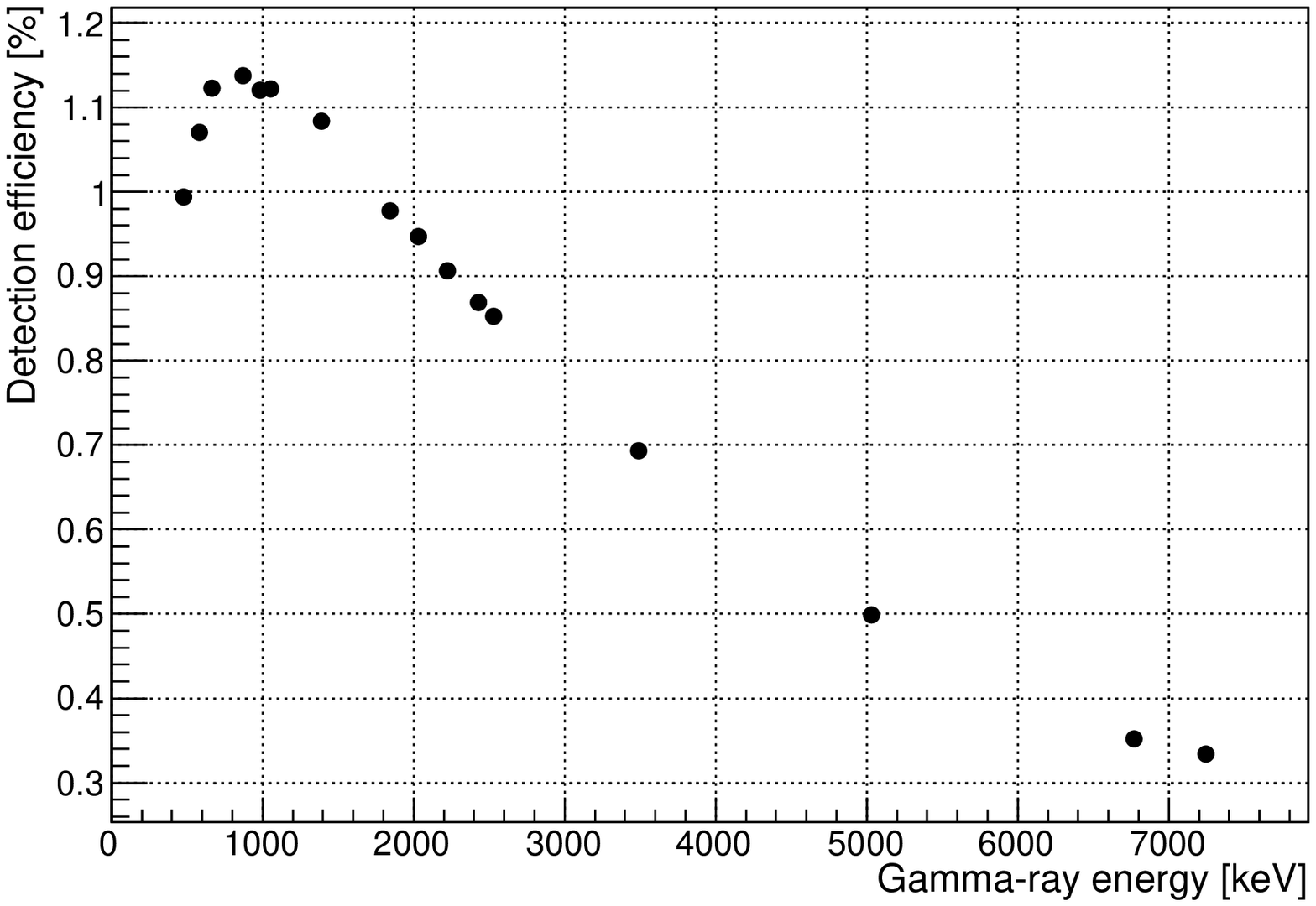}
		\caption{\label{fig:Efficiency} Energy dependence of the detection efficiency of full absorption peak.}
	\end{minipage}
	\end{tabular}
\end{figure}

We implemented the geometry of the germanium detector assembly by GEANT4 \cite{geant4}, and this simulation enabled us to calculate the response function of this detector units by emitting $\gamma$ rays from the target position. 
This simulation reproduced the measured spectra of radioactive sources of $^{137}$Cs and $^{152}$Eu in low energy regions.
The measured spectrum of the $^{14}$N(n, $\gamma$) reactions was used to verify the reproducibility of this simulation in a high energy region of up to about $11~{\rm MeV}$, and was reproduced by summing up the simulated spectra of monochromatic $\gamma$-ray energy that were scaled to confirm the intensity of the measured spectrum \cite{Takada}.
Figure~\ref{fig:Simuhist} shows an example of a simulated spectrum of monochromatic $\gamma$-ray with energy 7245~keV which was caused by the $^6$Li(n,$\gamma$) reactions, and an energy dependence of the detection efficiency of full absorption events is shown in figure~\ref{fig:Efficiency}. 

\begin{figure}[htbp]
\centering
\includegraphics[width=.7\textwidth]{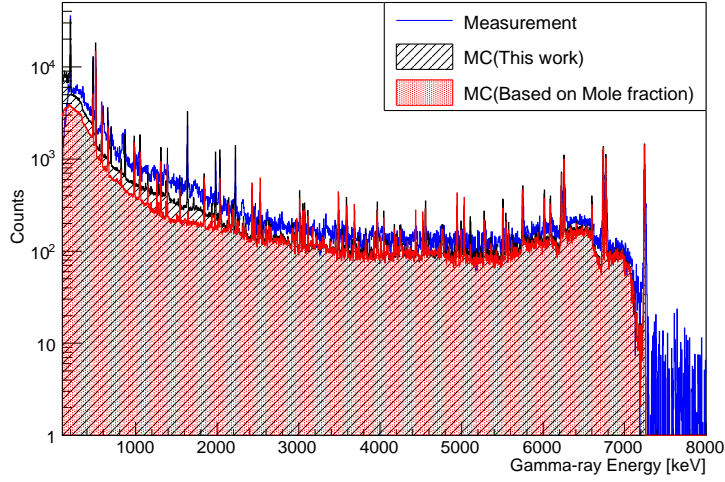}
\caption{\label{fig:Result} Comparison of the measured spectrum (solid line), the reconstructed spectrum (shaded area), and the spectrum based on the mole fraction of table~\ref{tab:LiF} (dotted area).}
\end{figure}

We estimated the intensity of each $\gamma$-ray peak from the higher energy region. 
Each simulated spectrum of monochromatic $\gamma$ rays was scaled to confirm the integrated value of the corresponding peak with three times of the Gaussian width, and removed from the measured spectrum. 
This procedure was repeated until all peaks were subtracted with higher energy.
In the $^{19}$F(n, $\gamma$) reactions, most of the peaks are too small to estimate from the measured spectrum.
Thus, the intensities of their peaks were calculated based on the reference values from the IAEA database \cite{IAEA}. 
These small peaks were scaled to confirm the relative intensities for the intense peak of $1381~{\rm keV}$ that is caused by the $^{19}$F(n, $\gamma$) reactions. 
Figure \ref{fig:Result} shows the measured spectrum, the reconstructed spectrum that sums up the all the scaled spectra of the monochromatic $\gamma$ rays, and the simulated spectrum which summed up the spectra of monochromatic $\gamma$-rays spectra scaled based on the mole fraction presented in table~\ref{tab:LiF}. 
In the low energy region at less than $2~{\rm MeV}$, the intensity of the reconstructed spectrum was larger than that of the simulated spectrum based on the mole fraction. 

Table \ref{tab:IntensityRatio} shows the amount of $\gamma$ rays emitted from the $^6$LiF tile in all directions by the (n, $\gamma$) reactions, and the ratios of the intensities measured in the experiment to those assumed from the mole fraction for the intense peaks. 
In the peak with $7245~{\rm keV}$, the ratio is normalized as unity, because this peak was initially scaled so that the intensity matched with the measured spectrum and the simulated spectrum in the reconstruction procedure.
The ratios that could not be compared with the assumed intensities for the unexpected peaks are described with horizontal lines in this table. 

It should be noted that there was an intense peak of 198~keV. 
According to the energy level of scheme, $\gamma$ rays with such energy would not be emitted by (n,$\gamma$) reactions of the elements that compose the $^6$LiF tile. 
In the neutron TOF spectrum of this peak, the decay time was $20.9 \pm 2.0~{\rm ms}$, and it was consistent with that of $20.4~{\rm ms}$ which is derived from the $^{71}{\rm Ge}({\rm n},\gamma)$ reaction. 
Therefore, the $\gamma$ rays with energy $198~{\rm keV}$ in this measurement are not influenced to the measurement of neutron lifetime, because these $\gamma$ rays were not emitted from the $^6$LiF tile.

\begin{table}[htbp]
\centering
\caption{\label{tab:IntensityRatio} The number of $\gamma$ rays emitted to all directions and the ratio of the intensity measured in the experiment to that assumed from the mole fraction for each intense peak. The errors described in this table are statistical errors.}
\smallskip
\begin{tabular}{|c|c|c|c|}
\hline
Energy [keV] & Element & Number of $\gamma$ rays ($\times 10^3$) & Ratio\\
\hline
7245 & $^6$Li & $1081 \pm 25.9$ & $1$ (fix)\\
6768 & $^6$Li & $1150 \pm 26.5$ & $1.71 \pm 0.06$\\
5033 & $^{19}$F & $47.5 \pm 3.16$ & $0.49 \pm 0.06$\\
3488 & $^{19}$F & $49.5 \pm 2.73$ & $0.44 \pm 0.03$\\
2529 & $^{19}$F & $56.8 \pm 2.65$ & $0.53 \pm 0.03$\\
2427, 2431 & $^{19}$F & $85.0 \pm 3.26$ & $0.85 \pm 0.04$\\
2223 & $^1$H & $273 \pm 6.20$ & - \\
2032 & $^7$Li & $209 \pm 5.17$ & $2.41 \pm 0.08$\\
1983 & - & $170 \pm 4.54$ & - \\
1843 & $^{19}$F & $74.8 \pm 2.87$ & $0.81 \pm 0.04$\\
1634 & - & $508 \pm 8.67$ & - \\
1387, 1392 & $^{19}$F & $155 \pm 4.07$ & $1.12 \pm 0.04$\\
1052 & $^7$Li, $^{19}$F & $257 \pm 4.79$ & $1.66 \pm 0.05$\\
980, 983 & $^7$Li, $^{19}$F & $226 \pm 4.49$ & $0.78 \pm 0.02$\\
869 & - & $2.02 \pm 1.35$ & -\\
656, 665 & $^{19}$F, $^{19}$F & $461 \pm 7.74$ & $0.87 \pm 0.03$\\
583 & $^{19}$F & $424 \pm 7.51$ & $0.78 \pm 0.02$\\
511 & e & $123 \pm 3.67$ & -\\
477 & $^6$Li & $2274 \pm 27.4$ & $3.9 \pm 0.09$\\
\hline
\end{tabular}
\end{table}

\subsection{Estimation of the neutron intensity and the number of $\gamma$ rays per neutron}
Our purpose is to know the intensity of prompt $\gamma$ rays from the $^6$LiF tile per an incident neutron. Therefore, we need to know the number of neutrons irradiated to the target. 
We calculated the neutron intensity based on the intense peak with energy $7245~{\rm keV}$. 
This peak has the highest-energy prompt $\gamma$ rays that are emitted from the $^6$LiF tile, and thus, it is suitable to estimate the neutron intensity due to few background events such as the Compton components of higher-energy $\gamma$ rays. 
The neutron intensity can be calculated by
\begin{align}
N &= \phi \epsilon \left\{1-\exp \left(- \rho \sigma \left(\frac{2200}{v} \right) B L_{\rm eff} \left(\frac{v}{2200}\right) \right) \right\} \nonumber \\
&\approx \phi \epsilon \rho \sigma B L_{\rm eff} \label{eq:flux},
\end{align}
where $N$ is an integrated value of the $7245~{\rm keV}$ peak with a range of three times of the Gaussian width, $\phi$ is the neutron beam intensity (the number of neutrons that are irradiated to the target), $\epsilon$ is the full absorption efficiency of the germanium detector units for $7245~{\rm keV}$ $\gamma$ rays, $\rho$ is the number density of $^6$Li, $\sigma$ is the neutron absorption cross-section for 2200~m/s neutrons, and $L_{\rm eff}$ is the effective thickness of the $^6$LiF tile. $B$ is the branching ratio of the $\gamma$ ray with energy 7245~keV emitted after neutron absorption. 
While $\sigma$ and $L_{\rm eff}$ depend on neutron velocity $v$ based on the $1/v$ law, and $v$ is canceled out, as shown in eq.~(\ref{eq:flux}).
Therefore, we can perform an analysis independent of TOF.
In the measured spectrum, $N$ was $(3.61 \pm 0.06) \times 10^3$. 
The full absorption efficiency $\epsilon$ was calculated from the simulation, and $\epsilon$ is $0.33\pm 0.03$\%. 
$\rho$ was $1.60 \times 10^{22}$ [$\#/{\rm cm^3}$] and $\sigma$ is $940 \pm 4$~barn, as described in table~\ref{tab:LiF}. 
$L_{\rm eff}$ was $6.5\times10^{-2}$~cm and $B$ is $2.63 \times 10^{-5}$ adopted from the IAEA database \cite{IAEA}.
The calculation results show that the number of incident neutrons $\phi$ was $(4.2 \pm 0.6) \times 10^{10}$ in the measurement with the $^6$LiF tile. 
We checked that this estimation was consistent with the measurement result using the $^{10}$B target described in appendix~\ref{app:BeamIntensity}.
 
Figure~\ref{fig:intensity} shows the intensity and the energy of prompt $\gamma$ rays emitted from the $^6$LiF tile per neutron, which was calculated using the number of incident neutrons as $(4.2 \pm 0.6) \times 10^{10}$. 
The intensity of each monochromatic $\gamma$ ray was calculated from each scale parameter which was used in the reconstruction. 

Table~\ref{tab:GammaRaysPerNeutron} summarizes the total number of the prompt $\gamma$ rays emitted from the $^6$LiF tile per incident neutron for each case. 
According to the IAEA database \cite{IAEA}, neutron capture cross section with $\gamma$ rays of observed impurities described in appendix~\ref{app:impurity} are larger than that of the constituent elements of the $^6$LiF tile. 
Considering the influence of these impurities, the number of $\gamma$ rays per neutron increases by about $1.05~{\rm times}$.
On the other hand, the number of total events of the reconstruct spectrum is 25\% less than that of the measured spectrum in figure~\ref{fig:Result}.
There are two possible explanations for this difference that have not been addressed in this study.
Firstly, the impurities in the $^6$LiF tile cannot be studied in detail.
If the elemental composition is different between the surface and the inside of this tile, the amount of the observed impurities may be different from that presented in table~\ref{tab:XRF}.
In addition, it may include other kinds of impurities that cannot be observed by the XRF measurement. 
Secondly, the simulation cannot estimate the effect of neutrons scattered by the target.
Scattered neutrons cause the $\gamma$ rays to be emitted from the other materials around the target, e.g., the detector assembly.
Such $\gamma$ rays are not likely to form full-absorption peaks.
The reason for this is that the solid angle and the detection efficiency for $\gamma$ rays from the surrounding materials is smaller than those from the target because the alignment of the germanium crystals is optimized to detect the $\gamma$ rays from the target position.
We cannot distinguish whether the difference of 25\% is caused by unexpected impurities or the effect of scattered neutrons.
Therefore, this difference was added asymmetrically to the systematic errors. 
The total number of the prompt $\gamma$ rays emitted from the $^6$LiF tile was $(2.3^{+0.7}_{-0.3} ) \times 10^{-4}$ per an incident neutron.
This is $1.4^{+0.5}_{-0.2}$ times the value assumed from the mole fraction in table~\ref{tab:LiF}.
The amount of $\gamma$ rays produced from the $^{6}{\rm LiF}$ tile is not more than twice the originally assumed value.
In the neutron lifetime measurement at J-PARC, the amount of background events produced in the TPC was 3.2 times of the originally expected value \cite{NeutronLifetimeJPARC}. 
However, the result of this study does not explain all of them.

\begin{figure}[htbp]
\centering
\includegraphics[width=.7\textwidth]{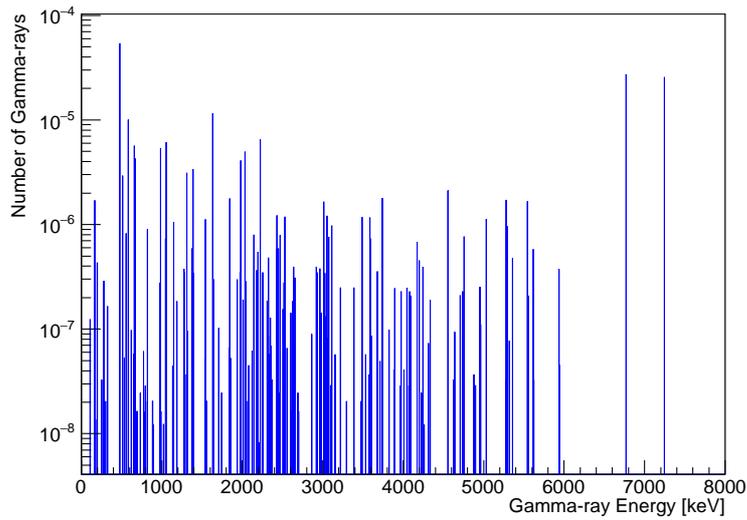}
\caption{\label{fig:intensity} The intensity of $\gamma$ rays with each energy emitted from the $^6$LiF tile per incident neutron.}
\end{figure}

\begin{table}[htbp]
\centering
\caption{\label{tab:GammaRaysPerNeutron} The number of total $\gamma$ rays per incident neutron ($\times 10^{-4}$) for each case. }
\smallskip
\begin{tabular}{cc}
\hline
Measurement result & $2.3^{+0.7}_{-0.3}$ \\
Mole fraction without impurities & $1.66 \pm 0.03$ \\
Mole fraction with impurities & $1.74^{+0.08}_{-0.04}$ \\
\hline
\end{tabular}
\end{table}


\section{Summary}
\label{sec:summary}
The neutron lifetime measurement experiment at J-PARC uses the $^6$LiF tile as the inner wall material in the detector to suppress background events.
The $\gamma$ rays emitted from the $^6$LiF tile cause the irreducible background events and a uncertainty for subtraction. 
The amount of the $\gamma$ rays must be measured to estimate the background events to realize precise measurement.
PGA measurement was conducted to estimate the number of prompt $\gamma$ rays emitted from the $^6$LiF tile per incident neutron. 
We obtained that this as $(2.3^{+0.7}_{-0.3}) \times 10^{-4}$, and this is $1.4^{+0.5}_{-0.2}$ times the value assumed from a mole fraction of the $^{6}{\rm LiF}$ tile.
We concluded that the amount of $\gamma$ rays produced from the $^{6}{\rm LiF}$ tile is not more than twice the originally assumed value.

\appendix
\section{Impurities in the $^6$LiF tile}
\label{app:impurity}
The impurities in the $^6$LiF tile were inspected by XRF measurements. In this measurement, the $^6$LiF tile with size $30 \times 30 \times 5~{\rm mm}^3$ was irradiated by X-ray in a chamber filled with He gas. The impurities were identified by measuring the XRF emitted by the de-exicitaion of electrons upon X-ray irradiations.
The result of XRF is shown in table~\ref{tab:XRF}.
\begin{table}[htbp]
\centering
\caption{\label{tab:XRF} Impurities in the $^6$LiF tile identified with XRF. N.D. refers to 'not detected'.}
\smallskip
\begin{tabular}{|c|c||c|c||c|c|}
\hline
Element & Content [\%] & Element & Content [\%] & Element & Content [\%]\\
\hline
F & Main material & Rb & N.D. & Tb & N.D.\\
Na & 0.04 & Sr & N.D. & Dy & N.D.\\
Mg & N.D. & Y & N.D. & Ho & N.D.\\
Al & 0.01 & Zr & N.D. & Er & N.D.\\
Si & 0.03 & Nb & N.D. & Tm & N.D.\\
\hline
P & 0.002 & Mo & N.D. & Yb & N.D.\\
S & 0.002 & Ru & N.D. & Lu & N.D.\\
Cl & N.D. & Rh & N.D. & Hf & N.D.\\
K & 0.004 & Pd & N.D. & Ta & N.D.\\
Ca & 0.007 & Ag & N.D. & W & N.D.\\
\hline
Sc & N.D. & Cd & N.D. & Re & N.D.\\
Ti & N.D. & In & N.D. & Os & N.D.\\
V & N.D. & Sn & N.D. & Ir & N.D.\\
Cr & N.D. & Sb & N.D. & Pt & N.D.\\
Mn & N.D. & Te & N.D. & Au & N.D.\\
\hline
Fe & N.D. & I & N.D. & Hg & N.D.\\
Co & N.D. & Cs & N.D. & Tl & N.D.\\
Ni & N.D. & Ba & N.D. & Pb & N.D.\\
Cu & N.D. & La & N.D. & Bi & N.D.\\
Zn & N.D. & Ce & N.D. & Th & N.D.\\
\hline
Ga & N.D. & Pr & N.D. & U & N.D.\\
Ge & N.D. & Nd & N.D. & - & -\\
As & N.D. & Sm & N.D. & - & -\\
Se & N.D. & Eu & N.D. & - & -\\
Br & N.D. & Gd & N.D. & - & -\\
\hline
\end{tabular}
\end{table}

\section{Measurement of beam intensity using $^{10}$B target}
\label{app:BeamIntensity}
\begin{figure}[htbp]
\centering
\includegraphics[width=.5\textwidth, bb=23.5 20 706 460]{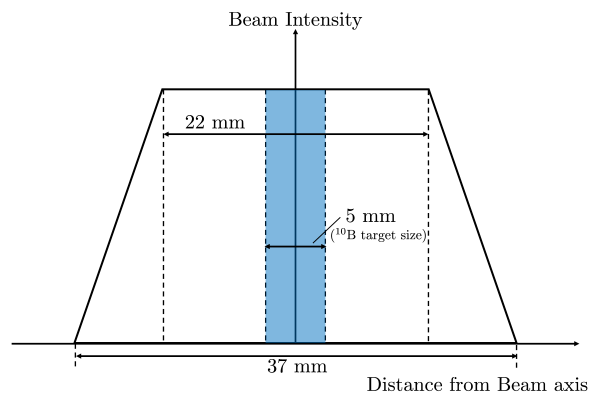}
\caption{\label{fig:Beamintensity} Spacial distribution of the neutron beam intensity. The intensity is constant within 22~mm in diameter and decreases linearly from 22~mm to 37~mm in diameter.}
\end{figure}

We measured the beam intensity using $^{10}{\rm B}$ of 5~mm diameter and 0.5~mm thickness in order to confirm the estimation of the neutron beam flux using the highest-energy peak. The measurement time was about 210 minutes.
Almost all neutrons irradiated onto this target were captured, and $\gamma$ rays with energy 477~keV were emitted via the $^{10}{\rm B}({\rm n}, \alpha)^{7}{\rm Li}^*$ reactions.
The number of detected $\gamma$ rays during the measurement with the $^{10}{\rm B}$ target is about $7 \times 10^7$, and thus, the number of 4$\pi$-emitted $\gamma$ rays was estimated about $7 \times 10^9$, because the detection efficiency for 477~keV $\gamma$ rays is about 1\%. 
Thus, the number of incident neutrons was estimated at about $7 \times 10^9$.
In order to compare the number of neutrons irradiated to the $^{10}{\rm B}$ target and the $^{6}{\rm Li}$ target, the size difference between these targets must be taken into account.
We assumed that the spatial distribution of the beam intensity in a circular truncated cone. 
Figure \ref{fig:Beamintensity} shows the spatial distribution of the beam intensity. 
It is constant within 22~mm in diameter, and decreases linearly from 22~mm to 37~mm in diameter.
The beam intensity for the $^{6}{\rm Li}$ target  corresponds to the volume of the circular truncated cone obtained by rotating the trapezoid ($V_1$), while that for the $^{10}{\rm B}$ target corresponds to the volume of the cylinder whose diameter with 5~mm ($V_2$).
The number of neutrons irradiated to the $^{6}{\rm Li}$ target can be estimated using the measurement result with the $^{10}{\rm B}$ target as follows:
\begin{align}
V_1 &= \pi \int_0^I \left( \frac{(37/2)-(22/2)}{I}x+\frac{22}{2} \right)^2 dx \sim 222.25 \pi I, \nonumber \\
V_2 &= \pi(5/2)^2 \times I \sim 6.25 \pi I, \nonumber \\
\phi &= (7 \times 10^9) \times \frac{35 \, {\rm min.}}{210 \, {\rm min.}} \times \frac{V_1}{V_2} \sim4.15 \times 10^{10},
\end{align}
where $I$ is a normalized neutron beam intensity at the center of the beam axis. 
This calculation result is consistent with the value estimated from the peak with energy 7245~keV.

\acknowledgments
The authors thank to the staff of ANNRI for the maintenance of the germanium detectors, and MLF and J-PARC for operating the accelerators. The neutron-irradiation experiment was performed under the user program (Proposal No.~2016A0245). This work was supported by the Neutron Scattering Program Advisory Committee of IMSS and KEK (Proposal No.~2014S03 and 2019S03). 


\end{document}